\documentstyle[12pt]{article}
\begin{document}
\begin{center}
\large
{\bf Anyon Zitterbewegung}
\end{center}
\vskip 2cm
Subir Ghosh\\
Physics Department,\\
Dinabandhu Andrews College,\\
Calcutta 700084, India.\\
\vskip 4cm
\noindent Abstract\\
A new model for anyon is proposed, which exhibits the classical analogue
of the quantum phenomenon - Zitterbewegung. The model is derived from
existing spinning particle model and retains the essential features of
anyon in the non-relativistic limit.
\vskip 4cm
PACS number(s): 11.10.Kk, 11.10.Ef, 03.20.+i, 11.10.Qr, 14.60.Cd
\newpage

In this Letter, we present a new Spinning Particle Model 
 (SPM) in 2+1-dimensions,
which is capable of producing a classical analogue of the well-known
quantum phenomenon - Zitterbewegung \cite{sch}. Recently it has been
demonstrated \cite{gkl}, in the context of existing SPM
of anyon \cite{jn, sg1}, that this oscillation in 2+1-dimensions is a
gauge artifact and can be gauged away with appropriate gauge fixing. Hence
this effect was thought to be reserved for 3+1-dimensional particles. We
will show that in a close variant of the established SPM this effect can
indeed show up, while the new model retains the essential anyonic
characteristics of the parent model, under certain plausible approximations.
Also the present scheme can pave the way for simulating various interactions.

Our starting model is relativistic, but for greater transparancy we have
gauge fixed the reparametrization invariance, thus losing manifest
relativistic covariance. The final results are obtained in the low energy
and heavy particle limit. We have closely followed the classic work of
Barut and Zanghi \cite{bz} in spirit. The analysis is carried through in
the Lagrangean framework.

It should be pointed out that after the original success of Chern-Simons
construction \cite{w} of anyon, theorists are more and more opting for
the SPMs of anyon \cite{pl}. Removal of the Chern-Simons statistical gauge
field and the induced "side effects" \cite{j} in the former option has
remained a long standing problem, whereas the economical and direct approach
in the latter framework \cite{pl,sg1} has obvious attractions.

Let us briefly describe the parent anyon model \cite{sg1},
\begin{equation}
{\cal L}=(M^2{\dot x}^{\mu}{\dot x}_{\mu}+{{J^2}\over 2}\sigma^{\mu\nu}
\sigma_{\mu\nu}+MJ\epsilon^{\mu\nu\lambda}{\dot x}_{\mu}\sigma_{\nu\lambda})
^{1\over 2},
\label{eql1}
\end{equation}
where $M$ and $J$ are the mass and (arbitrary) spin parameters and
$$ {\dot x}^{\mu}={{dx^{\mu}}\over{d\tau}};~~\sigma^{\mu\nu}=\Lambda
_{\lambda}^{~\mu}{{d\Lambda^{\lambda\nu}}\over{d\tau}};~~ \Lambda_{\lambda}
^{~\mu}\Lambda^{\lambda\nu}=\Lambda^{\mu}_{~\lambda}\Lambda^{\nu\lambda}=g^{\mu\nu};
~~g^{00}=-g^{11}=-g^{22}=1. $$
Hamiltonian analysis at once reveals that
$$(-{{\partial\cal L}\over{\partial\dot x_{\mu}}})^2\equiv p^{\mu}p_{\mu}=M^2;~~
(-{{\partial \cal L}\over{\partial \sigma_{\mu\nu}}})^2\equiv S^{\mu\nu}S_{\mu\nu}
=2J^2;~~ {1\over 2}\epsilon^{\mu\nu\lambda}S_{\mu\nu}p_{\lambda}=MJ,$$
where $S^{\mu\nu}$ is the relativistic spin. These relations ensure the
generic "rigid" dynamics, where spin and momentum vectors are directly
related \cite{sg1,pl}. We rewrite $\cal L$ to remove the square root,
\begin{equation}
{\cal L}={1\over{4\psi}}({\dot x}^{\mu} 
+{J\over{2M}}\epsilon^{\mu\nu\lambda}
\sigma_{\nu\lambda})
({\dot x}_{\mu} +{J\over{2M}}\epsilon_{\mu\alpha\beta}\sigma^{\alpha\beta})
+\psi M^2,
\label{eql2}
\end{equation}
and finally replace the inverse auxiliary field $\psi$,
\begin{equation}
{\cal L}=-(\dot x_\mu +{J\over{2M}}\epsilon_{\mu\alpha\beta}\sigma^{\alpha\beta})
A^\mu -(A^\mu A_\mu)\psi +M^2\psi,
\label{eql3}
\end{equation}
$A^\mu$ being another auxiliary field. From the equations of motion, we
rewrite $\cal L$ as 
\begin{equation}
{\cal L}=-\dot x^{\mu}p_{\mu}-{1\over 2}\sigma^{\mu\nu}S_{\mu\nu}-\psi (p^2-M^2),
\label{eql4}
\end{equation}
where we have renamed $p_\mu=-{{\partial {\cal L}}\over {\dot x^{\mu}}}=A_{\mu}$ 
and $S_{\mu\nu}=-{{\partial {\cal L}}\over
{\partial\sigma^{\mu\nu}}}={J\over M}\epsilon_{\mu\nu\lambda}A^{\lambda}$.

However, an equivalent and more convenient form of Lagrangean is the following
\cite{jn},
\begin{equation}
L=-\dot x^{\mu}p_{\mu}-\dot n^{\mu}P_{\mu}-{{\lambda}\over 2}(n^2+1)
-\lambda_1(p_\mu n^\mu)-\lambda_2(p_\mu P^\mu)-\Lambda (p^2-M^2).
\label{eqlag}
\end{equation}
Note that the canonically
conjugate spin variables $(n^\mu,~P_\nu)$, together with the
constraints coupled to the multipliers $\lambda$, $\lambda_1$ and $\lambda_2$,
are lumped into the previous non-canonical 
conjugate variables $(\sigma^{\mu\nu},~
S_{\mu\nu})$. The details are provided in \cite{sg1}.

It should be pointed out that two definitions of total angular momentum,
$$-\epsilon_{\mu\nu\lambda}(x^\nu p^\lambda+n^\nu P^\lambda )~ and~
-(\epsilon_{\mu\nu\lambda}x^{\nu} p^{\lambda} +J{{p_{\mu}}
\over{\sqrt{p^2}}})$$
are equivalent as far as angular momentum algebra and constraints are
concerned. However, due to the involved Dirac Bracket algebra, on the constraint
surface, a direct identification between $-\epsilon_{\mu\nu\lambda}n^\nu
P^{\lambda}$ and $-J{{p_{\mu}}\over{\sqrt{p^2}}}$ is not allowed 
with reference to the {\it full} gauge invariant sector \cite{bbg}.
The former relation is the natural choice for (\ref{eqlag}), whereas the
latter relation agrees with the relations derived from (\ref{eql1}).

The system of equations of motion and constraints resulting from (\ref{eqlag})
can be solved easily to yield,
\begin{equation}
\dot x^\mu =-2\Lambda p^\mu;~~\dot p^\mu =0;~~\dot n^\mu =0;~~\dot P^\mu
=\lambda n^\mu;~~\lambda_1=\lambda_2=0.
\label{eqmq}
\end{equation}
The multipliers $\lambda$ and $\Lambda$ corresponding to the first class
constraints remain undetermined. The natural choice of gauge, $x_0=\tau$
 (proper time) gives $\Lambda=-{1\over{2p_0}}$ and one obtains the solution
$x_i(\tau)={{p_i}\over{p_0}}\tau +~constant$, consistent with the free motion
with constant velocity \cite{jn}.

This analysis clearly shows that in this model, any oscillation in the $x$-
coordinate has to be trivial, since for the conventional parametrization
of $x_0$ as above, it is absent.

Our aim is to modify (\ref{eqlag}) judiciously such that in the same gauge as 
above, (ie. $x_0=\tau$), an extra oscillation is superimposed on the free
motion of $x_i$. But at the same time it is imperative to enforce the
constraints present in (\ref{eqlag}), so that the spinning particle
properties are preserved. The new Lagrangean is,
\begin{equation}
L_z=L-a\epsilon^{\mu\nu\lambda}P_\mu n_\nu p_\lambda,
\label{eqlz}
\end{equation}
where $a$ is a numerical parameter. 

The Lagrangean equations corresponding
to $(p,~x,~P,~n)$ are respectively,
$$\dot x^\mu =-2\Lambda p^\mu -\lambda_1 n^\mu -\lambda_2 P^\mu
-a\epsilon^{\mu\nu\lambda}P_\nu n_\lambda ,$$
$$\dot p^\mu =0 ,$$
$$\dot n^\mu =-\lambda_2 p^\mu -a\epsilon^{\mu\nu\lambda}n_\nu P_\lambda ,$$
\begin{equation}
\dot P^\mu =\lambda n^\mu +\lambda_1 p^\mu -a\epsilon^{\mu\nu\lambda}P_\nu
p_\lambda.
\label{eqmqz}
\end{equation}
Notice that the four constraints connected to $\Lambda,~\lambda,~\lambda_1$
and $\lambda_2$ remain unchanged. Using the equations of motion, time
persistance of the constraints on the constraint surface, determines
$\lambda_1=0$, (from ${{d(p^\mu P_\mu)}\over{d\tau}}=0$)
 and $\lambda_2=-{a\over{m^2}}
\epsilon^{\mu\nu\lambda}p_\mu n_\nu P_\lambda$, from 
(from ${{d(p_\mu n^\mu)}\over{d\tau}}=0$).
Time derivatives of the remaining constraints are identically satisfied.

Now we fix the gauge $x_0=\tau$ to determine $\Lambda$,
\begin{equation}
\Lambda=-{1\over{2p_0}}(1+\lambda_2 P_0+a\epsilon_{ij}P^in^j).
\label{eqLam}
\end{equation}
Lastly the gauge choice $P_0=~constant$ fixes $\lambda$,
\begin{equation}
\lambda={a\over{n_0}}\epsilon_{ij}P^ip^j.
\label{eqlam}
\end{equation}

Finally we restrict ourselves to the non-relativistic limit, ${{p_i}\over
M}\approx 0$ and $p_0\approx M$, which makes $\lambda_2\approx 0$. 
Incorporating all the above informations into (\ref{eqmqz}), we obtain,
$$\dot x^i\approx -2\Lambda p^i -an_0\epsilon^{ij}P_j;~~\dot P^i\approx
\lambda n^i -aM\epsilon^{ij}P_j,$$
\begin{equation}
\dot n_0\approx -a\epsilon^{ij}n_i P_j;~~\dot n^i\approx an_0\epsilon^{ij}P_j.
\label{eqnr}
\end{equation}
In the above set of equations, we can still drop the $\lambda$-term, (since
it contains $p_i$ whereas the other term in the $\dot P_i$ equation
has $M$), and reduce $\Lambda\approx -{1\over{2p_0}}$, (since in the $\dot x^i$
equation we neglect $\mid {{ap_i}\over M}\mid$ for small values of $a$). Thus
the two all-important equations, for our present purpose, are,
\begin{equation}
\dot x^i\approx -{{2p^i}\over M}-an_0\epsilon^{ij}P_j;~~\dot P^i\approx
-aM\epsilon^{ij}P_j.
\label{eqos}
\end{equation}
Clearly the behaviour of the spin variable $P_i$ is reminiscent of the
Barut-Zanghi construction \cite{bz}. Rewriting the $P_i$ equations in
terms of $P_{\pm}=P_i\pm iP_2$, we find $\ddot P_{\pm}=-(aM)^2P_{\pm}$,
which introduces an oscillatory term of frequency $aM$ in the $\dot x_i$
equation. Note that the frequency depends on $M$ in the correct way. 
This is the Zitterbewegung we were looking for.

There are other interesting applications in our proposed modification 
in (\ref{eqlz}). The $a$ term in (\ref{eqlz}) can be treated as an
interaction term by replacing one of the particle degrees of freedom
by an external c-number function field or by a suitable combination of
another particle's coordinates. The equations of motion will obviously
change. The advantage of the present scheme is that at every step one can 
adjust the parameters so that the fundamental anyonic behaviour is kept
intact as much as possible.

\end{document}